\documentclass[preprint,showpacs,preprintnumbers,amsmath,amssymb,prb]{revtex4}
\usepackage{graphicx}
\usepackage{dcolumn}
\usepackage{bm}
\usepackage{epsf}

\begin{document}

\title{Bloch-Siegert shift for multiphoton resonances}

\author{Peter Hagelstein}
\email{plh@mit.edu}
\author{Irfan Chaudhary}
\email{irfanc@mit.edu}

\affiliation{Research Laboratory of Electronics \\ 
Massachusetts Institute of Technology \\
Cambridge, MA 02139,USA}


\begin{abstract}
Recently there has been theoretical and experimental interest in Bloch-Siegert 
shifts in an intense photon field. 
A perturbative treatment becomes difficult in this multiphoton regime. 
We present a unitary transform and rotated model, which allows us to get accurate
results away from the level anticrossings.
A simple variational energy estimate leads to a new expression for the dressed
two-level system energy which is accurate, and useful over a wide
range of the dimensionless coupling constant.
\end{abstract}

\maketitle

\section{Introduction}

The basic problem of a two-level system coupled to an oscillatory perturbation 
arose in the early on in quantum mechanics in association with studies of spin dynamics 
in a magnetic field \cite{BlochSiegert,Shirley}.  
The closely related problem of a two-level coupled to a simple harmonic 
oscillator has been of interest for more than 30 years, at least since the work of Cohen-Tannoudji et al
\cite{Cohen}.
The Hamiltonian for this problem can be written as

\begin{equation}
\label{eq:BasicHamiltonian}
\hat{H} 
~=~
\Delta E {\hat{s}_z \over \hbar}
+
\hbar \omega_0 \hat{a}^\dagger \hat{a}
+
U (\hat{a}^\dagger+\hat{a}) {2 \hat{s}_x \over \hbar}
\end{equation}

\noindent
where we can write the spin operators $\hat{s}_i$ in terms of the Pauli matrices 
as 

\[
\hat{s}_i =  \frac{\hbar}{2} \hat{\sigma}_i
\]

\noindent
This model has been of interest recently in problems in which atoms interact 
with strong electromagnetic fields.
We have become interested in this problem since it can exhibit coherent energy 
coupling between systems with characteristic energies that are very different.
The multiphoton regime in which $\Delta E \gg \hbar \omega_0$ in this regard is 
of interest to us.
Additionally, our focus has been on the large $n$ limit.

It is well known that the dipolar interaction between photons and atoms causes 
the frequency of the atom to get shifted. 
The associated shift has been termed the Bloch-Siegert shift in the literature 
\cite{BlochSiegert, Shirley}.  
A general analytic expression for the $(2k + 1)$th resonance has been known for 
some time \cite{AhmadBullough}.  
Measurements of this shift have been reported recently up to $k = 11$ 
\cite{Fregenal}.  
Various theoretical methods have been used to understand these results 
\cite{OstrovskyHorsdal, Forre}.

We have recently found a unitary transformation that separates the problem into
three parts, each of which is relatively complicated.  In the
multiphoton regime we view the largest of
these as an ``unperturbed'' dressed Hamiltonian, which we have found give
energy eigenvalues that are close to those of the original problem away from
the level anticrossings.  In the present work, our attention is drawn to this 
aspect of the problem.  Using this approach, we are able to develop a reasonably good 
estimate of the Bloch-Siegert shift using a simple variational approximation, one which we 
can compare with results obtained using other methods.

\section{Rotated Hamiltonian}

As mentioned above, we have found it helpful to work in a rotated frame.
Instead of considering the Hamiltonian in Equation~(\ref{eq:BasicHamiltonian}), 
we consider the unitary equivalent Hamiltonian

\begin{equation}
\hat{H}^\prime 
~=~
e^{-i \hat{\lambda} \hat{\sigma}_y }
\hat{H}
e^{i \hat{\lambda} \hat{\sigma}_y }
\end{equation}

\noindent
where $\hat{\lambda}$ is 

\[
\hat{\lambda} ~=~ - \frac{1}{2} \arctan \left \lbrace
       {2 U (\hat{a} + \hat{a}^\dagger) \over \Delta E} \right \rbrace 
\]

\noindent
The operator $\hat{\lambda}$ depends explicitly on $\hat{a}+\hat{a}^\dagger$.

This rotation can be carried out explicitly, leading to a complicated 
expression.
It is useful to write the rotated Hamiltonian in the form

\begin{equation}
\hat{H}^\prime 
~=~
\hat{H}_0
+
\hat{V}
+
\hat{W}
\end{equation}

\noindent
where we have defined $\hat{H}_0$, $\hat{V}$, and $\hat{W}$ according to

\begin{equation}
\hat{H}_0
~=~
\sqrt{\Delta E^2 + 4 U^2 (\hat{a}+\hat{a}^\dagger)^2 } \left ( {\hat{s}_z \over 
\hbar} \right )
+
\hbar \omega_0 \hat{a}^\dagger \hat{a} 
\end{equation}

{\small

\begin{equation}
\hat{V} ~=~ 
{i \hbar \omega_0 \over 2}
\left \lbrace
\left [
{ \displaystyle{  U  \over \Delta E}  \over 1+  \left [ \displaystyle{ 2U 
(\hat{a}+\hat{a}^\dagger) \over \Delta E} \right ]^2  }
\right ]
(\hat{a}-\hat{a}^\dagger)
+
(\hat{a}-\hat{a}^\dagger)
\left [
{ \displaystyle{  U  \over \Delta E}  \over 1+   \left [ \displaystyle{ 2 U 
(\hat{a}+\hat{a}^\dagger) \over \Delta E} \right ]^2  }
\right ]
\right \rbrace
\left ( {2\hat{s}_y \over \hbar} \right )
\end{equation}
}

{\small

\begin{equation}
\hat{W} ~=~ \hbar \omega_0 
\left \lbrace
{ \displaystyle{  U  \over \Delta E}  \over 1+   \left [ \displaystyle{
      2 U (\hat{a}+\hat{a}^\dagger) \over \Delta E} \right ]^2  } 
\right \rbrace^2
\end{equation}

}

Even though this rotated Hamiltonian is much more complicated than the one we started out with,  
we have found that it has useful properties that are of interest.  
For example, from numerical calculations and also from the WKB approximation, 
we have found that the ``unperturbed'' part of the Hamiltonian $\hat{H}_0$ is 
a good approximation to the initial Hamiltonian $\hat{H}$ away from level
anticrossings.  
The ``perturbation'' $\hat{V}$ contains the part of the Hamiltonian that is most
responsible for level splitting at the anti-crossing (as we will discuss 
elsewhere).
Finally, the remaining term $\hat{W}$ gives rise to a minor correction to 
$\hat{H}_0$.

\section{Bloch-Siegert shift}

To illustrate the usefulness of this rotation, we focus on one issue in 
particular: the
development of an interesting approximation for the energy eigenvalues.
For this, we need focus only on $\hat{H}_0$.
Consider the time-independent Schr\"odinger equation for the ``unperturbed'' 
part of
the rotated problem

\begin{equation}
E \psi ~=~ \hat{H}_0 \psi
\end{equation}

\noindent
Because $\hat{H}_0$ is somewhat complicated, it is not straightforward to 
develop exact
analytic solutions.  We have used numerical solutions to develop some 
understanding of 
the eigenfunctions and associated eigenvalues.  From these numerical studies, it 
has
become clear that there exists a useful analytic approximation.  

We adopt an approximate solution of the form

\begin{equation}
\psi_t ~=~ |n \rangle |s,m \rangle
\end{equation}  

\noindent
where $|n \rangle$ is a pure SHO eigenstate, and $|s,m \rangle$ is a spin 
function.  
The variational estimate for the energy eigenvalue in this approximation is

\begin{equation}
E_t ~=~ \langle \psi_t | \hat{H}_0 | \psi_t \rangle
\end{equation}

\noindent
This approximate energy can be evaluated explicitly to give

\begin{equation}
E_t 
~=~ 
\bigg \langle n \bigg | 
\sqrt{\Delta E^2 + 4 U^2 (\hat{a}+\hat{a}^\dagger)^2 } 
\bigg | n \bigg \rangle m 
~+~ 
n \hbar \omega_0
\end{equation}

\noindent
This result is interesting in that it leads directly to a simple interpretation
of the system in terms of a dressed two-level system and unperturbed oscillator,
which is reasonably good away from the level anticrossings.  Moreover, the 
dressed
two-level system energy is involved in the calculation of the Bloch-Siegert
shift in the literature.  

Let us write the variational energy as

\begin{equation}
E_t ~=~ \Delta E(g)m + \hbar \omega_0n
\end{equation}

\noindent
where we define $g$ according to

$$g ~=~ {U \sqrt{n} \over \Delta E}$$

\noindent
and $\Delta E(g)$ as

\begin{equation}
\Delta E(g) ~=~
\Delta E
\bigg \langle n \bigg | 
\sqrt{1 + {4 g^2 (\hat{a}+\hat{a}^\dagger)^2 \over n} } 
\bigg | n \bigg \rangle 
\label{approximation}
\end{equation}

\begin{figure} [t]
\begin{center}
\mbox{\epsfbox{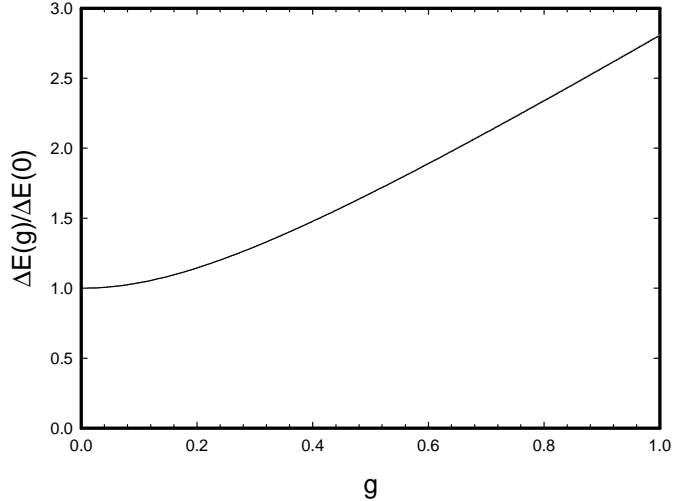}}
\caption{Plot of $\Delta E(g)$ computed from direct numerical solutions for 
$\Delta E=11 \hbar \omega_0$
  and $n=10^8$, and also the approximate $\Delta E(g)$ based on the 
approximation of Equation~(\ref{approximation}).
  The two curves are so close that they cannot be distinguished.} 
\label{Fig32x1}
\end{center}
\end{figure}

\noindent
This approximation is very good, as can be seen in Figure~\ref{Fig32x1}
where the approximation is plotted along with the exact numerical results. 
The Bloch-Siegert resonance condition can be written in terms of $\Delta E(g)$ as 

\begin{equation}
\label{eq:ResonanceCondition}
\Delta E(g) ~=~ (2k + 1)  \hbar \omega_0
\end{equation}

\epsfxsize = 4.00in
\epsfysize = 2.50in

It is possible to develop a simpler approximation for this approximate dressed 
energy based on the WKB approximation.  We may write

\begin{equation}
\Delta E(g)
~=~
{\Delta E \over \pi}
\int_{-\sqrt{\epsilon}}^{\sqrt{\epsilon}}
\sqrt{
1 + 8 g^2 y^2/n 
\over 
\epsilon - y^2
}
dy
\end{equation}

\noindent
with $\epsilon = 2n+1$.  Results from this approximation can also not be distinguished from
the exact results if plotted as done previously in Figure 1.

\section{Comparison with other results}

As the results available in the literature of interest here are given for small 
$g$, we require
a power series expansion of Equation (\ref{approximation}):

\begin{equation}
\label{eq:SeriesExpansion}
\bigg \langle n \bigg | 
\sqrt{\Delta E^2 + 4 U^2 (\hat{a}+\hat{a}^\dagger)^2 } 
\bigg | n \bigg \rangle
~\longrightarrow~
\Delta E
\left [
1 
+
4 g^2
-
12
g^4
+
\cdots
\right ]
\end{equation}

To compare to the exact expression obtained by Ahmad and
Bullough\cite{AhmadBullough}, we write the
resonance condition using their notation as 

\begin{equation}
\frac{\omega_0^2}{\omega^2} 
~=~ 
(2n+1)^2 
\left[ 1 - \frac{2}{n(n+1)}
\frac{b^2}{\omega^2} +\frac{n^2 + n - 1}{2n^3(n+1)^3}
\frac{b^4}{\omega^4} + \cdots 
\right ]
\end{equation}

\noindent
We translate this into our notation to obtain 

\begin{equation}
\frac{\Delta E^2}{\omega_0^2} 
~=~ 
(2k +1)^2 
\left [
1 - 
\frac{2}{k(k+1)}
\frac{U^2 n}{\omega_0^2} +\frac{k^2 + k - 1}{2k^3(k+1)^3}
\frac{U^4 n^2}{\omega_0^4} + \cdots 
\right ]
\end{equation}

\noindent
Keeping in mind the resonance condition [Equation (\ref{eq:ResonanceCondition})], 
this is consistent with

\begin{equation}
\label{eq:AhmadBullough}
\Delta E(g) 
~=~
\Delta E
\bigg [ 1 + a(k) g^2 + b(k) g^4 + \cdots \bigg ]
\end{equation}

\noindent
where

$$a(k) ~=~ 4 + {1 \over k(k+1)}$$

$$b(k) ~=~ -12 - { 8k^4 + 16 k^3 + 3k^2 - 5k -1 \over 4 k^3 (k+1)^3}$$
\noindent We can now see that in the large $k$ limit,
Equation~(\ref{eq:AhmadBullough}) reduces to our result. It should also be
noted that the deviations for finite $k$ are $O(1/k^2)$.

Another approximate multiphoton result is given by Ostrovsky and
Horsdal-Pedersen\cite{OstrovskyHorsdal}.  
This result (written in their notation) for the Bloch-Siegert shift is 

\begin{equation}
N \omega ~=~ \omega_{ba} \left(1 + \frac{1}{4} q^2 - \frac{3}{64}q^4 +
O(q^6)  \right)
\end{equation}

\noindent
This result is equivalent to our result [Equation (\ref{eq:SeriesExpansion})] to fourth order in $g$.

\section{Conclusion}

The unitary equivalent Hamiltonian provides a simple way to understand the Bloch-Siegert shift. 
We have found in our work that the unperturbed part of the rotated Hamiltonian 
($\hat{H}_0$) produces a reasonably accurate estimate of the dressed two-level system energy 
in the large $n$ limit away from the level anticrossings.  
In the discussion presented here, we have shown that a simple variational estimate for the energy 
associated with $\hat{H}_0$ in the rotated frame produces good results away from the level anticrossings
over a wide range of the dimensionless coupling strength $g$.
The approximation is found to be in agreement with previous work in the limit that $g$ is small.

\appendix

\end{document}